\documentclass{article}
\usepackage{amsfonts}
\usepackage{epsfig,amsmath}
\usepackage{natbib}

\newcommand{\ex}[1]{\ensuremath{\mathbb{E}[#1]}}

\newcommand{\corr}[1]{\ensuremath{\mathrm{Corr}[#1]}}
\newcommand{\bd}[1]{\ensuremath{\mbox{\boldmath $#1$}}}

\begin{document}

\title{Locally adaptive spatial smoothing using conditional autoregressive models}

\author{Duncan Lee $^{1}$ and Richard Mitchell $^{2}$\\
$^{1}$School of Mathematics and Statistics, University of Glasgow, Glasgow, UK\\
$^{2}$Institute for Health and Wellbeing, University of Glasgow, Glasgow, UK}


\maketitle


\begin{abstract}
{Conditional autoregressive (CAR) models are commonly used to capture spatial correlation in areal unit data, and are typically specified as a prior distribution for a set of random effects, as part of a hierarchical Bayesian model. The spatial correlation structure induced by these models is determined by geographical adjacency, so that two areas have correlated random effects if they share a common border. However, this correlation structure is too simplistic for real data, which are instead likely to include sub-regions of strong correlation as well as locations at which the response exhibits a step-change. Therefore this paper proposes an extension to CAR priors, which can capture such localised spatial correlation. The proposed approach takes the form of an iterative algorithm, which sequentially updates the spatial correlation structure in the data as well as estimating the remaining model parameters. The efficacy of the approach is assessed by simulation, and its utility is illustrated in a disease mapping context, using data on respiratory disease risk in Greater Glasgow, Scotland.}

{Conditional Autoregressive priors; Disease mapping; Integrated Nested Laplace Approximations; Spatial correlation}
\end{abstract}

\section{Introduction}
Data arising from non-overlapping spatial units are prevalent in many fields, including agriculture (\cite{besag1999}), ecology (\cite{brewer2007}), education (\cite{wall2004}), epidemiology (\cite{lee2011}) and image analysis (\cite{gavin1997}).  The set of areal units can form a regular lattice or differ largely in both shape and size, with examples of the latter including a set of electoral wards or census tracts corresponding to a city or county. In either case such data typically exhibit spatial correlation, with observations from areal units close together tending to have similar values. This spatial correlation is caused by the existence of spatial structure in the covariate risk factors that affect the response, some of which are unknown or unmeasured. This unmeasured spatial structure is typically modelled by a set of random effects in a hierarchical Bayesian model, which are assigned a  conditional autoregressive (CAR, \cite{besag1974}) prior. If no covariates are available to capture the spatial pattern in the response then the random effects represent the overall spatial pattern in the response, where as otherwise they capture any structure in the residuals. CAR models are most often specified as a set of univariate conditional distributions, where the conditioning is only on the values of the random effects in areal units that are geographically close. This geographical information is contained in a neighbourhood or weight matrix $W$, which summarises the level of spatial similarity between each pair of neighbouring areas. Typically, a binary value of one or zero is specified for each $w_{kj}$, depending on whether areal units $(k,j)$ share a common border. Thus, if areas $(k,j)$ share a common border ($w_{kj}=1$) their random effects are correlated, where as otherwise ($w_{kj}=0$) they are conditionally independent. A number of CAR models have been proposed in the literature, including the intrinsic and BYM models (both \cite{besag1991}), as well as alternatives  developed by \cite{leroux1999} and \cite{stern1999}.\\  

However, these priors model the level of spatial correlation globally across the entire region being studied, which is likely to be overly restrictive for real data. This is because spatial data are likely to contain areas of smooth evolution in the response variable, as well as locations at which abrupt step-changes occur. An example is when modelling the spatial pattern in health, deprivation or educational attainment across a city, where very different communities, such as those that are rich and poor, often live side by side. In this context the different communities are likely to have very different values of the quantity being modelled, and are hence unlikely to be correlated even though they share a common border. Thus, the value of the response variable is likely to be correlated within each community, but not across the border where the two communities meet. A motivating example is shown in Figure \ref{figure SIR}, which displays the risk of respiratory disease across Greater Glasgow, Scotland. The figure highlights that respiratory disease risk is far from being spatially smooth, with many high risk areas (the darker shaded regions) being  geographically adjacent to much lower risk areas (the lighter shaded regions). Such step-changes are known as \emph{boundaries} in the response surface, and their identification are of direct interest, in addition to suggesting that existing global spatial correlation models are inappropriate. This is because their locations may reflect changes in the underlying biological, physical or social processes underlying the response (\cite{jacquez2000}),  and may represent the border between two different communities or neighbourhoods.\\

A number of approaches have been proposed for modelling localised spatial structure and identifying discontinuities in areal unit data, including the use of local statistics (\cite{boots2001}), mixture models (\cite{knorrheld2000}), and extending the class of CAR priors (\cite{lu2005}, \cite{lu2007}, \cite{li2011} and \cite{lee2012}). In this paper we follow the latter approach, and model localised spatial correlation by estimating the elements of the neighbourhood matrix $w_{kj}$ as one or zero as long as areas $(k,j)$ share a common border, rather than assuming they are fixed at one. This is because $w_{kj}$ determines the partial correlation between the random effects in areas $(k,j)$ (conditional on the remaining random effects), and if $w_{kj}=1$ they are correlated, while if $w_{kj}=0$ they are conditionally independent. Thus, if $w_{kj}=1$ the random effects are smoothed over in the modelling process, where as if $w_{kj}=0$ they are not. The approach proposed in this paper is an iterative algorithm, which iterates between estimating $W$ and the model parameters conditional on the other until a termination criterion is met. This approach is made practical in a computational sense by using Integrated Nested Laplace Approximations (INLA, \cite{rue2009}) to estimate the model parameters, which is very much faster than Markov Chain Monte Carlo (MCMC) simulation.\\

The motivating application for this methodological development is the area of disease mapping, whose aim is to  quantify the spatial pattern in disease risk over a city or country, so that areas with high risks can be identified. The production of such maps  play an important role in public health improvement, as they enable clusters of high risk areas to be identified. This in turn allows appropriate remedial action to be taken, such as targeting an educational campaign about the risk factors, or investigating whether there are any harmful exposures in that area. The remainder of this paper is organised as follows. Section 2 provides the background  to areal unit data modelling, including the specification of a general hierarchical model, as well as a review of existing approaches to localised spatial smoothing in this context. Section 3 presents our proposed methodological extension, which focuses on capturing the localised spatial structure in the data via estimation of the neighbourhood matrix $W$. Section 4 assesses the efficacy of our approach using simulation, while Section 5 identifies boundaries in the risk surface for respiratory disease cases in Greater Glasgow, Scotland. Finally, Section 6 contains a concluding discussion and outlines future developments.

\section{Background}

\subsection{Spatial models for areal unit data}
Consider a study region consisting of $n$ non-overlapping areal units $\mathcal{S}=\{A_{1},\ldots,A_{n}\}$, on which a set of responses $\mathbf{Y}=(Y_{1},\ldots,Y_{n})$, and a vector of known offsets $\mathbf{O}=(O_{1},\ldots,O_{n})$, are observed. The spatial pattern in the response is modelled by a matrix of $p$ covariates $X=(\mathbf{x}_{1}^{\tiny\mbox{T}\normalsize},\ldots,\mathbf{x}_{n}^{\tiny\mbox{T}\normalsize})$ and  a set of random effects $\bd{\phi}=(\phi_{1},\ldots,\phi_{n})$, the latter of which are included to model any spatial correlation that remains in the data after the covariate effects have been accounted for. The vector of covariates for areal unit $k$ are denoted by $\mathbf{x}_{k}^{\tiny\mbox{T}\normalsize}=(1, x_{k2},\ldots,x_{kp})$, the first of which corresponds to an intercept term. A general Bayesian hierarchical model for these data is given by

\begin{eqnarray}
Y_{k}|\mu_{k}&\sim&f(y_{k}|\mu_{k},\sigma)~~~~\mbox{for }k=1,\ldots,n,\nonumber\\
g(\mu_{k})&=&\mathbf{x}_{k}^{\tiny\mbox{T}\normalsize}\bd{\beta} + \phi_{k} + O_{k},\label{equation general model}\\
\phi_{k}|\bd{\phi}_{-k}, W,\tau,\rho&\sim&\mbox{N}\left(\frac{\rho\sum_{i=1}^{n}w_{ki}\phi_{i}}{\rho\sum_{i=1}^{n}w_{ki} + 1-\rho}~,~\frac{1}{\tau(\rho\sum_{i=1}^{n}w_{ki} + 1-\rho)}\right),\nonumber\\
\beta_{j}&\sim&\mbox{N}(0,1000)~~~~\mbox{for }j=1,\ldots,p,\nonumber\\
\tau,\sigma&\sim&\mbox{Gamma}(0.001, 0.001),\nonumber\\
\mbox{logit}(\rho)&\sim&\mbox{N}(0, 100).\nonumber
\end{eqnarray}

The response data $Y_{k}$ come from an exponential family of distributions $f(Y_{k}|\mu_{k}, \sigma)$, which includes the binomial, Gaussian and Poisson families as special cases. The expected value of $Y_{k}$ is denoted by $\ex{Y_{k}}=\mu_{k}$, while $\sigma$ is an additional precision parameter if required (for example if $Y_{k}$ is Gaussian). The expected value of the response is related to the linear predictor via an invertible link function $g(.)$, and common examples include the logit (binomial responses), identity (Gaussian responses) and natural log (Poisson responses) functions. Relatively diffuse priors are specified for the regression parameters $\beta_{j}$ and the precision parameters $\tau$ (and $\sigma$ if necessary), because typically there is little prior information available about their values. The diffuse Gaussian prior for $\rho$ on the logit scale is specified because it is the default choice when fitting the model above using INLA, the inferential approach adopted here.\\

The random effects $\bd{\phi}$ are assigned a prior from the class of conditional autoregressive models, which are a type of Gaussian Markov Random Field (GMRF). A number of priors have been proposed for modelling spatial correlation within this class, including the convolution model (\cite{besag1991}), as well as the alternative proposed by \cite{leroux1999}. The latter is used here because a comparative study by \cite{lee2011} showed that it was the most appealing in practice. CAR priors are most often written in terms of $n$ univariate full conditional distributions, and the Leroux model is written in that form in equation (\ref{equation general model}) above. The conditioning in this model is on the values of the random effects in geographically adjacent areas, the information on which is contained in the binary neighbourhood matrix $W$. The conditional expectation is a weighted average of the random effects in neighbouring areas and a global mean of zero, because the numerator is equivalent to $\rho\sum_{j=1}^{n}w_{kj}\phi_{j} + (1-\rho)\times 0$. The parameter $\rho$ determines the global level of spatial correlation between the random effects, with $\rho=0$ corresponding to independence everywhere, while $\rho$ equal to one defines strong spatial correlation throughout the region (simplifies to the intrinsic CAR prior). The set of full conditionals for all $n$ random effects is equivalent to the joint distribution, $\bd{\phi}\sim\mbox{N}(\mathbf{0},[\tau Q(\rho, W)]^{-1})$, where $Q(\rho, W)$ is a sparse precision matrix that is invertible if $\rho\in[0,1)$. It is given by

\begin{equation}
Q(\rho,W)~=~[\rho(\mbox{diag}(w_{k+}) - W) + (1-\rho)I],\label{equation lerouxjoint}
\end{equation}

where $\mbox{diag}(w_{k+})$ is a diagonal matrix with elements equal to the rowsums of $W$, while $I$ is an $n\times n$ identify matrix. The partial correlation between $(\phi_k, \phi_j)$  conditional on the remaining random effects is given by

\begin{equation}
\corr{\phi_{k},\phi_{j}|\bd{\phi}_{-kj}}~=~\frac{\rho w_{kj}}{\sqrt{(\rho\sum_{i=1}^{n}w_{ki} + 1-\rho)(\rho\sum_{i=1}^{n}w_{ji} + 1-\rho)}}\label{equation partialcorrelation}.
\end{equation}

Thus, for non-neighbouring areas where $w_{kj}=0$,  $(\phi_{k},\phi_{j})$ are conditionally independent given the values of the remaining random effects. In contrast, for neighbouring areas where $w_{kj}=1$, the random effects are correlated. This suggests that if there is substantial spatial correlation between the majority of pairs of random effects, i.e. if $\rho$ is estimated as close to one, then all pairs of random effects that share a common border will be spatially correlated. This model therefore enforces global spatial smoothing, and cannot capture both localised spatial smoothness and discontinuities in the random effects surface. Furthermore, the strength of the partial correlations depend on the number of neighbouring areas (through $\sum_{i=1}^{n}w_{ji}$), rather than on the underlying strength of the relationship between $(\phi_{k}, \phi_{j})$. These characteristics are undesirable, and in the next section we extend the class of CAR models to addresses these deficiencies, thus enabling areas of smooth evolution and step changes to be modelled in the random effects surface. However, a number of extensions to CAR priors have already  been proposed to address this problem, and before we outline our proposed solution we provide a critique of this existing work.

\subsection{Existing work on localised spatial smoothing}
In the model described above the joint posterior distribution being estimated is $f(\bd{\Theta}|\mathbf{Y},W)$, the posterior distribution of the model parameters $\bd{\Theta}=(\bd{\beta}, \bd{\phi}, \rho, \tau, \sigma))$ conditional on the data and a fixed neighbourhood matrix $W$ based on geographical adjacency. As previously discussed this is overly restrictive, and a more flexible approach would be to estimate the joint posterior distribution of the model parameters and the unknown spatial structure, i.e. $f(\bd{\Theta},W|\mathbf{Y})$. A small number of approaches have been proposed in this vein, including \cite{lu2007}, who treat  the set of $\{w_{kj}\}$  as binary random quantities if areas $(k,j)$ share a common border, rather than assuming they are fixed at one. This allows $(\phi_{k},\phi_{j})$ to be conditionally independent ($w_{kj}=0$) or correlated ($w_{kj}=0$), depending on the value of $w_{kj}$. In this context if $w_{kj}=0$ then a boundary is said to exist between the two random effects, as they are conditionally independent despite sharing a common border. The set of $\{w_{kj}\}$ are modelled using logistic regression, while similar approaches have been proposed by \cite{ma2007}, and \cite{ma2010}, who use a CAR prior and an Ising model respectively. However, this approach introduces a large number of extra parameters into the model, for example, in the study region considered in Section 5 there are $n=271$ data points and 701 individual elements $\{w_{kj}\}$. Furthermore these are covariance parameters, and in the related field of geostatistics it is well known that estimating two parameters (the range and smoothness of the spatial correlation) in the Matern correlation model  leads to identifiability issues. Therefore, treating $W$ as a matrix of additional unknown parameters results in a highly overparameterised covariance model for $\bd{\phi}$, and \cite{li2011}  suggest that individual $\{w_{kj}\}$ are poorly identified from the data and are computationally expensive to update.\\

Therefore an intermediate approach is required, which allows the elements of $W$ to be estimated, but without treating them as individual random quantities in an extra level of the Bayesian hierarchical model. In this vein \cite{lee2012} proposed modelling them using covariates and a small number of parameters, i.e. $W=W(\bd{\alpha})$ for parameters $\bd{\alpha}$, so as to produce a parsimonious model for the covariance structure of the random effects. However, their approach is crucially dependent on the existence of relevant covariate information to describe the spatial structure amongst the random effects, which may not always be available. An alternative approach was proposed by \cite{li2011}, who  consider different $W$ matrices as different models, and use the Bayesian Information Criterion (BIC) to choose between them. However, the number of possible models is $2^{\mathbf{1}^TW\mathbf{1}/2}$, and to get around the large model space they only consider models that have one boundary, as part of a data mining technique.

\section{Methods}

\subsection{Model and estimation}
The overall model is given by equation (\ref{equation general model}), and the unknown quantities being estimated are  the model parameters $\bd{\Theta}=(\bd{\beta}, \bd{\phi}, \rho, \tau, \sigma)$ and the binary elements of $W$. Only the elements of $W$ corresponding to areas sharing a common border are updated, with all other elements being fixed at zero. In this model $\rho$ controls the amount of spatial smoothing (correlation) globally across the study region, while as can be seen from (\ref{equation partialcorrelation}), $w_{kj}$ specifically determines whether $(\phi_k, \phi_j)$ are smoothed over (correlated) at this global level or not. In the case that $w_{kj}$ is estimated as zero a boundary in the random effects surface is said to have been identified between areas $(k,j)$. However, we note that if $\rho$ is estimated as close to zero then the boundary interpretation of $w_{kj}=0$ is lost, because $(\phi_k, \phi_j)$ will be approximately conditionally independent even if $w_{kj}=1$. Therefore, depending on the goal of the analysis it may be desirable to fix $\rho$ at a value close to one to enforce strong global smoothing, which can then be altered locally by  estimating the elements of $W$. A further discussion of this point is given in Section 3.2 below.\\

We propose an iterative estimation algorithm for $(\bd{\Theta}, W)$, which cycles between updating $\bd{\Theta}$ and $W$ conditional on the other until a termination criterion is met.  Conditional on $W$ the estimation of $\bd{\Theta}$ is fully Bayesian, with the joint posterior distribution being obtained from model (\ref{equation general model}) using INLA. In contrast, the elements of $W$ are treated as hyperparameters, which are deterministically updated conditional on the posterior distribution of $\bd{\Theta}$. Thus, the elements of $W$ are not updated in a fully Bayesian setting, because only estimates are provided rather than full posterior distributions. A fully Bayesian approach is not adopted for the reasons discussed above, namely  because it leads to a highly over-parameterised covariance model for the random effects. The next three subsections describe the estimation of $\bd{\Theta}|W$, $W|\bd{\Theta}$, and present the overall iterative algorithm. A set of functions to implement binomial, Gaussian and Poisson specifications of the model described below in the statistical package  \texttt{R}, (\cite{R2009}) are provided in the supplementary material accompanying this paper.

\subsubsection{Estimation of $\bd{\Theta}|W$}
The standard approach for estimating the parameters in a Bayesian model is MCMC simulation, which can be implemented in generic software such as WinBUGS (\cite{winbugs}). However, the estimation algorithm proposed here requires sequentially fitting model (\ref{equation general model}) with different but fixed neighbourhood matrices $W$, which would make an MCMC approach computationally impractical if the number of different $W$ matrices considered is large. Therefore, we propose estimating $\bd{\Theta}$ using Integrated Nested Laplace Approximations, the speed of which makes this approach computationally inexpensive. We note that the use of INLA in an areal unit context is relatively new, and existing examples are given by \cite{schrodle2011} and \cite{schrodle2011b}.

\subsubsection{Estimation of $W|\bd{\Theta}$}
We propose a deterministic approach for estimating the elements of the hyperparameter matrix $W$ conditional on the joint posterior distribution of $\bd{\Theta}$, which alleviates the identifiability problems associated with treating $\{w_{kj}\}$ as a set of binary random quantities. For adjacent areas $(k,j)$, $w_{kj}$ determines whether the random effects $(\phi_{k},\phi_{j})$ are correlated ($w_{kj}=1$) or conditionally independent ($w_{kj}=0$), and its value is updated by considering the current posterior distributions of $(\phi_{k},\phi_{j})$. If the 95$\%$ credible intervals of $\phi_{k}$ and $\phi_{j}$ do not overlap there is evidence that the two random effects are substantially different, and $w_{kj}$ is set equal to zero. In contrast, if the intervals do overlap then there is no substantial difference between the two random effects, and $w_{kj}$ is set equal to one. This approach thus induces spatial smoothing between the next estimates of $(\phi_{k},\phi_{j})$ if the current estimates are similar, where as no such smoothing is enforced if the current estimates are substantially different.

\subsubsection{Overall algorithm}
The iterative estimation algorithm consists of the following steps.\\

\begin{center}
\parbox{13cm}{
\emph{Algorithm}

\begin{description}
\item[1:] Estimate a starting joint posterior distribution for the model parameters using INLA, which is  denoted by $f(\bd{\Theta}^{(0)}|\mathbf{Y}, W^{(0)})$. For this initial model we assume the random effects are independent, which is achieved by restricting model (\ref{equation general model}) by fixing $\rho=0$.

\item[2:] Iterate the following two steps for $i=1,2,\ldots,i^{*}$, until one of the two termination conditions  for the hyperparameter matrix $W$, outlined in step 3, is met.

\begin{description}
\item[\textbf{a:}]  Estimate $W^{(i)}$ deterministically from the current joint posterior distribution $f(\bd{\Theta}^{(i-1)}|\mathbf{Y}, W^{(i-1)})$. Set $w^{(i)}_{kj}=1$ if the 95$\%$ credible intervals for $(\phi_{k}^{(i-1)},\phi_{j}^{(i-1)})$ overlap and areas $(k,j)$ share a common border. Otherwise, set $w^{(i)}_{kj}=0$.

\item[\textbf{b:}]  Estimate the joint posterior distribution $f(\bd{\Theta}^{(i)}|\mathbf{Y},W^{(i)})$ of model (\ref{equation general model}) using INLA. 
\end{description}

\item[\textbf{3:}] After $i^*$ iterations one of the following two termination conditions will apply.

\begin{description}
\item[\textbf{case 1}] - The sequence of $W$ estimates is such that $W^{(i^*)}=W^{(i^*+1)}$, which is the estimated hyperparameter matrix $\hat{W}$.

\item[\textbf{case 2}] - The sequence of $W$ estimates forms a  cycle of $k$ different states $(W^{(i^*)}, W^{(i^*+1)}, \ldots,W^{(i^*+k-1)}, W^{(i^*+k)})$, where $W^{(i^*)}=W^{(i^*+k)}$.  In this case the estimated hyperparameter matrix $\hat{W}$ is the value from the cycle of $k$ states that has the minimal level of residual spatial correlation, as measured by the absolute value of Moran's I statistic, (\cite{moran1950}), a measure of spatial autocorrelation.
\end{description}

When one of the termination conditions has been met $\hat{W}$ is the estimated spatial structure of the random effects, and $\bd{\Theta}$ is summarised by the joint posterior distribution $f(\bd{\Theta}|\mathbf{Y},\hat{W})$.
\end{description}}
\end{center}

The algorithm is initialised by assuming the random effects are independent (obtained by fixing $\rho=0$, in which case $f(\bd{\Theta}^{(0)}|\mathbf{Y}, W^{(0)})=f(\bd{\Theta}^{(0)}|\mathbf{Y})$), because this does not enforce any initial spatial smoothing constraints on the random effects surface. One of the two termination conditions outlined in the algorithm are guaranteed to apply after a sufficiently large number of iterations $i^{*}$, because each $w_{kj}$ is binary and hence the sample space for $W$ is large but finite (equal to $2^{\mathbf{1}^TW\mathbf{1}/2}$).  In practice (see section 4.4), the algorithm terminates by reaching a steady state (i.e. case 1 above where $W^{(i^*)}=W^{(i^*+1)}$) in almost all cases in a small number of steps, and case 2 only accounts for a small number of data sets.

\subsection{Interpretation and usage}
The model defined above can be applied in two main ways,  the most appropriate of which will depend on the goal of the analysis.  If the goal is to explain the spatial variation in the response, then any important covariates should be included in the mean model, and the random effects will capture any localised residual spatial structure, due to the existence of unmeasured confounders. In contrast, if the goal of the analysis is to identify the number and locations of any boundaries in the response surface, then covariates should be excluded from the mean model, because this ensures that $\bd{\phi}$ and $\bd{\mu}$ have the same spatial structure (as $\mu_k=g^{-1}(\beta_0 + \phi_k)$). This means that any boundaries identified in the random effects surface are also boundaries in the risk surface. In addition, $\rho$ should be fixed close to one (such as 0.99) in model (\ref{equation general model}), so that the random effects are modelled as being spatially smooth except where $w_{kj}=0$. This is required because if $\rho$ is estimated as zero, then  (\ref{equation general model}) and (\ref{equation partialcorrelation}) show that $\{w_{kj}\}$ would disappear from the model and hence do not affect whether two random effects are partially correlated or conditionally independent. Finally, we note that we would not fix $\rho=1$ in this case, because this results in the precision matrix $Q(1,W)$ being singular. Furthermore, if an area is not estimated to have any neighbours, i.e. if $\sum_{i=1}^{n}w_{ki}=0$ for some area $k$, then the full conditional distribution given by (\ref{equation general model})  has an undefined mean and variance (due to dividing by zero).

\section{Simulation study}
In this section we present a simulation study, that compares the performance of the adaptive local spatial smoothing model proposed in the previous section, against the commonly used alternatives. The latter include the BYM model (\cite{besag1991}) and the model proposed by \cite{leroux1999}, which both enforce a global level of spatial smoothness on the random effects surface. For this study the simulated response data are counts and a Poisson log-linear specification of model (\ref{equation general model}) is used, because that is the data type modelled in the disease mapping application in Section 5.

\subsection{Data generation}
We base our study on the $n=271$ areas that comprise the Greater Glasgow and Clyde health board, which is the region considered in the case study presented in Section 5. Simulated responses are generated from a Poisson log-linear specification of model (\ref{equation general model}), where for simplicity, an offset term is not included. The linear predictor includes an intercept term, a single covariate and a set of spatially correlated random effects, and the intercept term and the coefficient of the covariate are fixed for all simulations at $\ln(40)$ and 0.1 respectively. A new realisation of the random effects and the single covariate are generated for each simulated data set,  because this ensures the results are not affected by a particular realisation of either quantity. \\

\begin{figure}
\centering\caption{Locations of the boundaries (bold black lines) in the simulated random effects surfaces.}
\label{figure boundaries}\scalebox{0.5}{\includegraphics{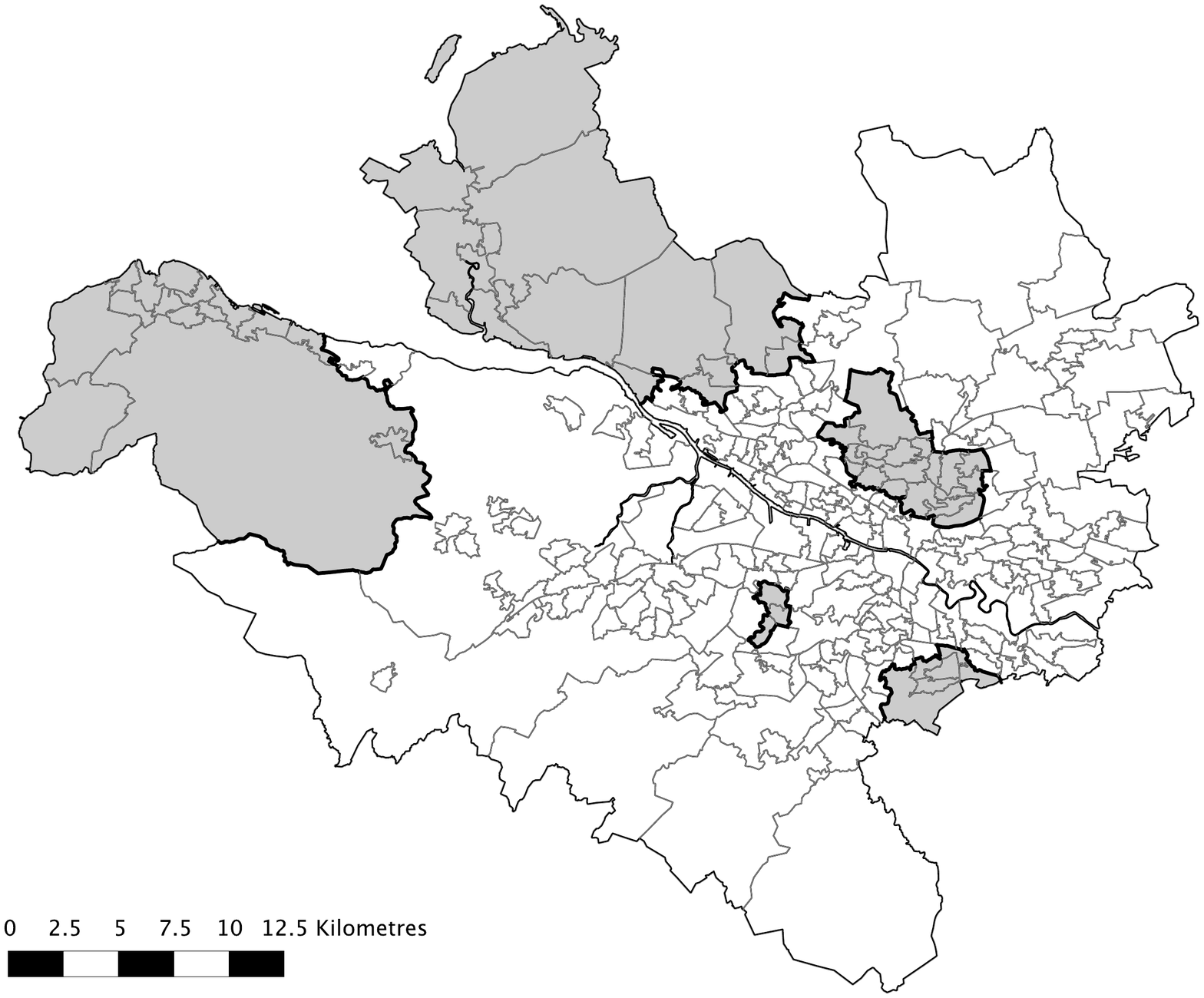}}
\end{figure}

The covariate is generated from a Gaussian distribution with a mean of zero and a variance of one, while the random effects are  generated from a multivariate Gaussian distribution. The variance matrix for the latter is specified by the spatially smooth Matern class of correlation functions with smoothness parameter equal to 2.5 and a spatial range of 5 kilometres, the latter of which is chosen so that the average correlation between pairs of areas is 0.5. The mean vector for $\bd{\phi}$ is piecewise constant, which enables us to induce localised spatial smoothness and boundaries in the random effects surface. This localised structure follows the template shown in Figure \ref{figure boundaries}, which partitions the study region into 6 groups, the main area shaded in white and the 5 smaller clusters shaded in grey. The mean of the random effects is zero for areas in the white region and $m$ for areas in the grey regions, which induces boundaries where the two regions meet (the bold black lines in Figure \ref{figure boundaries}). In this study we consider two scenarios, Scenario A - $m=1$ and Scenario B - $m=0$, the former corresponds to step changes in the random effects surface, while the latter relates to a spatially smooth surface with no step changes. When $m=1$ there are 74 boundaries in total across the study region, which is approximately 10$\%$ of the total number of borders. The results of the study are presented in two parts, the first quantifies the estimation performance of the three modelling approaches, while the second assesses the accuracy of the localised spatial structure identified by the locally adaptive model.

\subsection{Results - estimating fitted values and regression parameters}
Five hundred simulated data sets are generated for each of scenarios A and B, and the results are shown in Table \ref{table simulation}. The table displays the bias and root mean square error (RMSE) of both the fitted values $\bd{\mu}=(\mu_{1},\ldots,\mu_{n})$ and the single regression parameter $\beta$, and all  are presented as percentages of their true values. In addition, the table also displays the coverage probability of the 95$\%$ credible interval for the regression parameter $\beta$, which is again presented on the percentage scale. The table shows that all models produce close to unbiased estimates of the fitted values and the regression parameter in both scenarios, with percentage biases being less than $1\%$ in all cases. In contrast, the RMSE values are lower for the locally adaptive model in the presence of steps changes (scenario A), with reductions of around 5$\%$ ($\bd{\mu}$) and 8$\%$ ($\beta$) compared with the two global smoothing models. The results from scenario B suggest that in the absence of localised spatial correlation all models perform comparably, with biases and RMSE values being similar in all cases. Finally, while the inclusion of localised spatial structure into the random effects does adversely affect the estimation performance of the global correlation models (via the RMSE values), it does not affect the credible intervals for $\beta$, as all intervals have close to their nominal 95$\%$ coverage.\\

\begin{table}
\caption{\label{table simulation}Summary of the simulation study results. Scenario A represents random effects with step changes, while in Scenario B they are spatially smooth.} 
\centering\begin{tabular}{llrr}
\hline
\raisebox{-1.5ex}[0pt]{\textbf{Metric}}  &\raisebox{-1.5ex}[0pt]{\textbf{Model}} &\multicolumn{2}{c}{\textbf{Results}}\\
&&\textbf{Scenario A ($m=1$)}&\textbf{Scenario B ($m=0$)}\\\hline

&BYM&-0.782&-0.553\\
$\%$ Bias - $\bd{\mu}$&Leroux&-0.782&-0.535\\
&Locally adaptive&-0.355&-0.522\\\hline

&BYM&14.658&8.718\\
$\%$ RMSE  - $\bd{\mu}$&Leroux&14.674&8.647\\
&Locally adaptive&9.602&8.584\\\hline

&BYM&-0.146&0.541\\
$\%$ Bias - $\beta$&Leroux&-0.156&0.533\\
&Locally adaptive&-0.438&0.463\\\hline

&BYM&20.487&13.316\\
$\%$ RMSE - $\beta$&Leroux&20.592&13.296\\
&Locally adaptive&12.860&13.349\\\hline

&BYM&94.8$\%$&96.6$\%$\\
Coverage probability - $\beta$&Leroux&94.8$\%$&96.8$\%$\\
&Locally adaptive&95.6$\%$&96.4$\%$\\\hline
\end{tabular}
\end{table}

\subsection{Results - estimating localised spatial smoothness and boundaries}
The second part of this study assesses whether the locally adaptive model can identify the locations of the boundaries (step-changes) in the random effects surface, via its iterative estimation of the neighbourhood matrix $W$. For this part of the study the single covariate is removed from the data generation, so that the random effects have the same spatial structure as the fitted values (so that the step changes apply to both surfaces). In addition, as discussed in Section  3.2 $\rho$ is fixed equal to 0.99, so that the random effects are modelled as spatially smooth except where $w_{kj}$ is estimated as zero. As before, 500 data sets are generated under both scenarios A and B, and we consider two metrics to quantify the accuracy of the model. The first is boundary agreement (BA), which is the percentage of true boundaries, i.e. the bold black lines in Figure \ref{figure boundaries},  that are correctly identified as such by the model. The second metric is the  Non-Boundary Agreement (NBA), which measures the percentage of non-boundaries that the model correctly identifies. We note that these two metrics quantify the accuracy with which $W$ is estimated by the model. In scenario A the boundary agreement is 97.1$\%$ and non-boundary agreement is 99.7$\%$, suggesting that the model can correctly identify the locations of both boundaries and non-boundaries in the random effects (and fitted values) surface. In scenario B only NBA can be calculated, because as $m=0$ the random effects surface is spatially smooth and there are no true boundaries to identify. In this scenario NBA=99.7$\%$, which again suggests that the model does not identify large numbers of false positive, i.e, boundaries that are not real.

\subsection{Results - algorithm convergence}
The locally adaptive spatial smoothing model has been applied to 2000 simulated data sets in this study, which allows us to quantify empirically  the termination properties of the estimation algorithm proposed in section 3.1.3. The algorithm terminated under case 1 (i.e. $W^{(i^*)}=W^{(i^*+1)}$) for 1999 of the 2000 data sets (99.95$\%$), taking between 1 and 9 iterations to do so. For the remaining one data set the algorithm terminated under case 2, where the length of the cycle was only 2 different $W$ matrices.

\section{Case study}
We illustrate our methodology in the area of disease mapping, by modelling the spatial pattern in respiratory disease risk in Greater Glasgow, Scotland, in 2005. We utilise our model to address two different epidemiological questions: (i) what is the spatial pattern in respiratory disease risk and what factors affect it? and (ii) where and how many boundaries are there in the estimated risk surface? 

\subsection{Data description}
\begin{figure}
\centering\caption{Standardised Incidence Ratio (SIR) for respiratory disease.}
\label{figure SIR}\scalebox{0.5}{\includegraphics{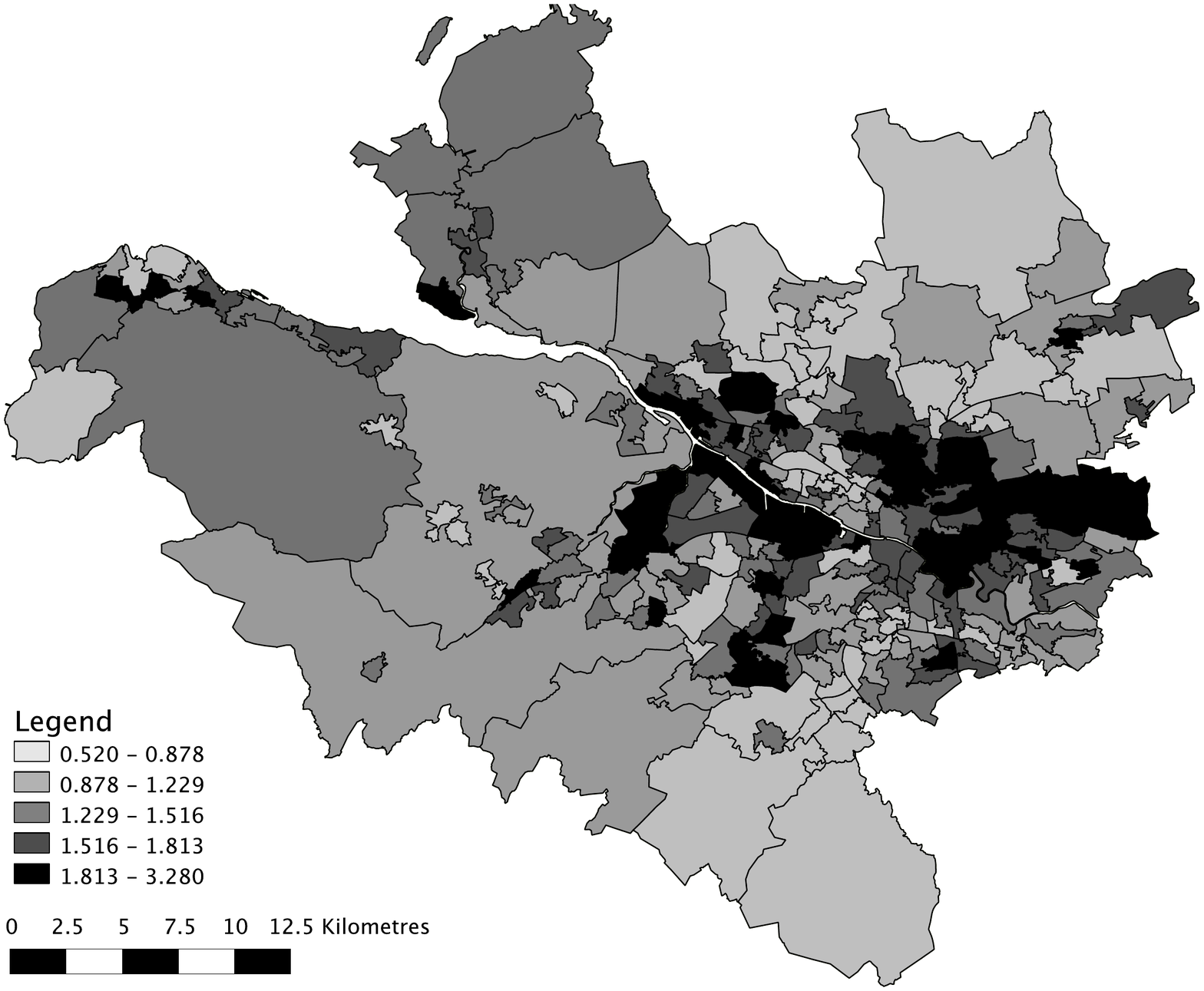}}
\end{figure}

The study region is the Greater Glasgow and Clyde health board, which contains the city of Glasgow in the east, and the river Clyde estuary in the west. Glasgow is the largest city in Scotland, with a population of around 600,000 people. The study region is partitioned into $n=271$ Intermediate Geographies (IG), which have a median population of 4,239 residents. All the data used in this study are publicly available, and can be downloaded from the Scottish Neighbourhood Statistics (SNS) database (\emph{http://www.sns.gov.uk}). The disease data comprise the numbers of people admitted to hospital with a main or secondary diagnosis of respiratory disease during 2005, which corresponds to ICD-10 codes J00-J99 and R09.1. The expected numbers of cases (the natural log of which is used as an offset in the model) were calculated by external standardisation, using age and sex specific rates for the whole of Scotland. The standardised incidence ratio (SIR) is the most informative scale on which to display these raw data, and is simply the ratio of the observed divided by the expected numbers of cases. Figure \ref{figure SIR}, displays the SIR for the 271 IGs across Greater Glasgow, which shows that the risks are highest in the heavily deprived east end of Glasgow (east of the study region), as well as along the southern bank of the river Clyde.\\

We consider two potential covariates in this study, which have previously been shown to affect respiratory ill health. The first of these is a measure of ethnicity, because different rates of respiratory disease have been observed for people from different ethnic backgrounds (\cite{LAIA2001}). The only variable available to measure ethnicity is the percentage of school children in each IG from ethnic minorities (non-white), which we appreciate is imperfect in many ways (for example, it does not differentiate between different ethnic groups).  The second variable we consider is a measure of socio-economic deprivation (\cite{holgate2007}), specifically the percentage of people in each IG that are defined to be income deprived (are in receipt of a combination of means-tested benefits). Finally, we note that a modelled estimate of the percentage of people in each IG that smoke is also available, although it is highly correlated with the income deprivation variable (Pearson's r=0.90). Both variables were included in separate Bayesian models (fitted using INLA) in conjunction with the ethnicity variable (without random effects), and as income deprivation produced the model with the lowest Deviance Information Criterion (DIC, \cite{spiegelhalter2002}), 2477.6 compared with 2626.5, it is used in this study rather than smoking.

\subsection{Modelling}
As the respiratory admissions data are a set of counts, a Poisson log-linear specification of model (\ref{equation general model}) is used in this section. The expected value of the response is given by $\mu_{k}=E_{k}R_{k}$, where $E_{k}$ denotes the expected number of cases in area $k$, while $R_{k}$ represents the overall disease risk in that area. Combining this with a natural log link function gives the general model $\ln(\mu_{k})=\log(E_{k})+\log(R_{k})=\log(E_{k})+\mathbf{x}_{k}^{\tiny\mbox{T}\normalsize}\bd{\beta} + \phi_{k}$. Initially, a Bayesian model with both covariates but no random effects was fitted to the data using INLA, to determine whether the random effects were required. However, the ethnicity covariate is highly skewed to the right, as the majority of the intermediate geographies have a very small (or zero) percentage of people who are non-white. Therefore a second model with a log transform of this covariate (a constant of 0.5  was added to prevent the occurrence of $\ln(0)$) was also fitted to the data, and as this model has a lower DIC than the previous one (2477.6 compared with 2488.3), the log transformation was retained. \\

The adequacy of the covariate only model was then assessed, and substantial overdispersion was found (overdispersion parameter equal to 3.28), as well as spatial correlation in the residuals. The latter was assessed via a Moran's I permutation test, which provided a p-value for the null hypothesis of no spatial correlation of 0.0080. To alleviate these problems a set of random effects were added to the model, and we compare the locally adaptive smoothing model proposed here, against the BYM model and the alternative proposed by Leroux given in (\ref{equation general model}). The goodness-of-fit of these models is summarised in Table \ref{table results}, which displays their DIC values. The table shows that the locally adaptive smoothing model proposed in Section 3 fits the data better than the global models, having a lower DIC value by around 14.

\subsection{Results}

\subsubsection{Covariate effects}
The covariate effects are shown in Table \ref{table results}, which displays posterior medians as well as 95$\%$ credible intervals. All results are presented on the relative risk scale, for a one standard deviation increase in each covariates value. There is strong evidence that income deprivation affects respiratory ill health, with populations exhibiting 14$\%$ higher levels of income deprivation having between  a 33.5$\%$ and a 35.1$\%$ increased risk of respiratory disease. In contrast, the statistical importance of the ethnicity effect is unclear, as the credible intervals are on the borderline of including the null risk of one.

\begin{table}
\caption{\label{table results} Summary of the models.}
\centering
\begin{tabular}{lrrr}
\hline \raisebox{-0.5cm}{\textbf{Model}}&\raisebox{-0.5cm}{\textbf{DIC}}&\multicolumn{2}{c}{\textbf{Covariate effects}}\\
&&\textbf{Income deprivation}& \textbf{Ethnicity}\\\hline

BYM&2065.9&1.351      (1.313,      1.390) &0.982      (0.952,      1.014) \\
Leroux&2067.5&1.350      (1.313,	1.389) &0.972      (0.944, 		1.000)\\
Locally adaptive &2051.5&1.335  (1.296,  1.375)& 0.967  (0.935,  0.999)\\
\hline
\end{tabular}
\end{table}

\subsubsection{Risk maps}
\begin{figure}
\centering\caption{Maps displaying the estimated spatial pattern in respiratory disease risk. The top panel displays the estimate from a model with covariates and random effects, while the bottom panel shows the results from a model with only random effects, which allows boundaries (thick white lines) to be identified.}\label{figure results}
\begin{picture}(20,20.5)
\put(0,2){\scalebox{0.45}{\includegraphics{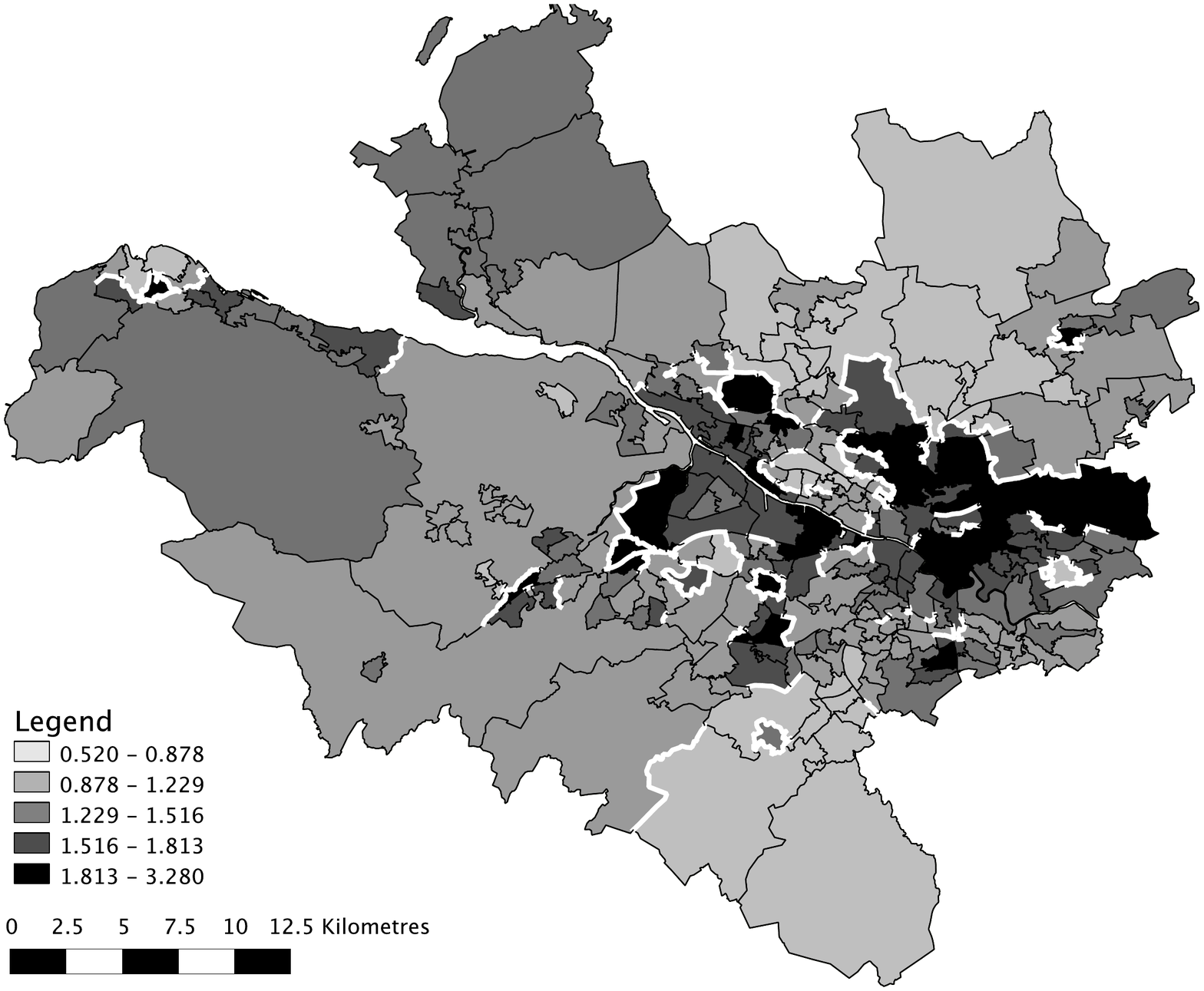}}}
\put(0,11){\scalebox{0.45}{\includegraphics{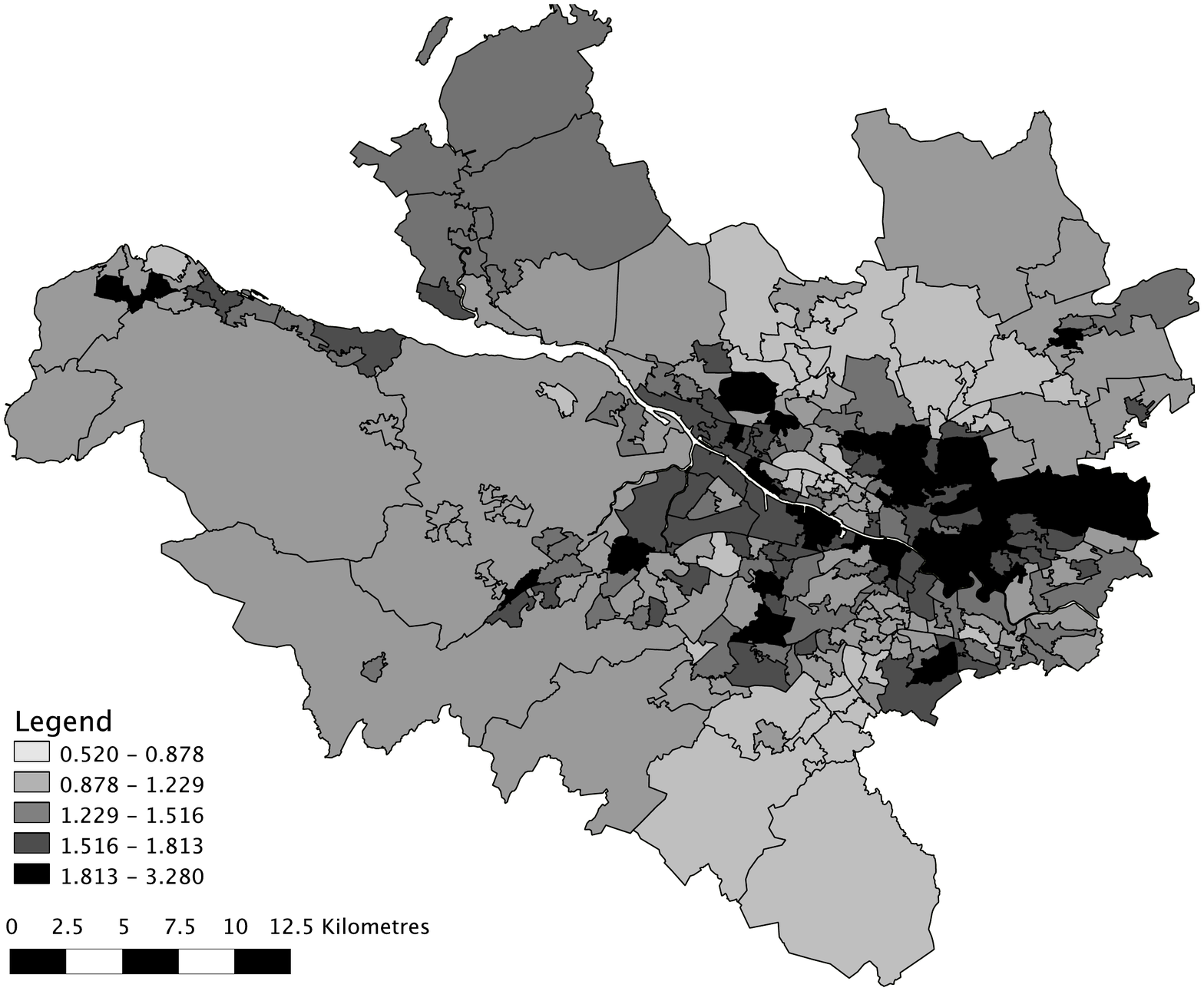}}}
\end{picture}
\end{figure}

The top panel of Figure \ref{figure results} displays the spatial pattern in respiratory disease risk estimated from the model described above, which includes both the covariates and the random effects. The quantity being displayed 
is disease risk, which is computed as $R_{k}=\exp(\mathbf{x}_{k}^{\tiny\mbox{T}\normalsize}\bd{\beta} + \phi_{k})$ and summarised by its posterior median. The risks are relative to the expected numbers of admissions, which were calculated using rates from the whole of Scotland rather than Greater Glasgow alone. Risks greater than one correspond to unhealthy areas relative to Scotland overall, while values less than one are low risk areas. The mean risk across Greater Glasgow is 1.407, suggesting that on average Greater Glasgow has a 40.7$\%$ greater risk of respiratory disease than Scotland overall. The highest risk areas are predominantly in the heavily deprived east-end of the city of Glasgow (east of the map), while the lowest risk areas are in the west end of the city (centre of the map) and in the outlying suburbs.\\

The bottom panel in Figure \ref{figure results} displays the spatial pattern in respiratory disease risk estimated from a model with only the random effects, because as discussed in Section 3.2, it allows boundaries in the risk surface to be identified. The risk surface is broadly similar to that estimated from the model including covariates, and the white lines denote the locations of risk boundaries, which separate areas that are geographically adjacent but have very different risks. There are 139 boundaries in total, which comprises just under 20$\%$ of the total number of boundaries in the study region. This suggests that respiratory disease risk in Greater Glasgow is far from being spatially smooth, and the boundaries identified appear to correspond to sizeable changes in the estimated risk surface. The model has identified two different types of boundaries. The first are a small number of closed boundaries, which enclose an area (or areas) that is (are) different from all of its (their) neighbours. The second type of boundaries are not closed, and simply separate two areas or groups of areas that have very different risks. While closed boundaries may be more appealing than non-closed ones, for example because the set of areas are partitioned into disjoint clusters, they may be too restrictive to apply in all cases. An illustrative example of this is the short boundary in the north-east of the study region, which partially surrounds a single high-risk area. This area has four geographical neighbours, three of which are low risk and hence are separated by a boundary, while one of which has a medium risk and hence is not separated by a boundary. Finally, we note that areas on opposite banks of the river Clyde (the thin white line running south east) are not assumed to be neighbours, which explains the absence of boundaries in this area.

\section{Concluding discussion}
This paper has proposed a novel approach for modelling localised spatial correlation and discontinuities in area unit data using CAR priors, by estimating the elements of the neighbourhood $W$ rather than assuming they are fixed based on geographical adjacency. An iterative estimation algorithm has been proposed, which adaptively alters the neighbourhood matrix depending on the posterior distributions of the remaining model parameters. The advantages of this approach are that it is fast to implement (the analysis in section five can be implemented in less than a minute), is fully automatic in terms of boundary identification, and does not require additional covariate information to identify the boundaries. In addition, it is easy to implement using the set of functions provided in the supplementary material accompanying this paper, and together, these facets are missing from the existing approaches proposed by \cite{lu2005}, \cite{lu2007} and \cite{lee2012}. The disadvantage of this approach is that it does not provide a fully Bayesian treatment of the elements of $W$, as only estimates are provided rather than full posterior distributions. As a result, the uncertainty in $W$ is not accounted for when estimating the model parameters. Instead, the elements of $W$ are treated as hyperaparameters, which are iteratively updated conditional on the remaining model parameters until a termination condition is met. The reasons for this is that $W$ contains a large number of covariance parameters, much larger than the number of data points, and \cite{li2011} have argued that such parameters are computationally expensive to update in a fully Bayesian context and the resulting posterior distributions are only weakly informative.\\

The simulation study has shown that in the presence of localised spatial structure the model proposed here outperforms the global spatial smoothing models, in terms of estimating both covariate effects and the overall mean response. The improvement in estimation is  in terms of RMSE, and ranges between 5$\%$ and 8$\%$ in the simulated data considered here. The simulation study thus emphasises the importance of correctly controlling for unmeasured confounding when estimating covariate effects, as specifying an inappropriate correlation structure results in poorer estimation performance in terms of RMSE. When the residual spatial structure is completely smooth with no step changes the locally adaptive model performs as well as the global models, which taken together with the previous results suggests that it could be considered the method of choice in this context. The study also shows that the algorithm can accurately identify the locations of the step changes in the response (estimate the elements of $W$), with accuracy rates all upwards of 97$\%$ for the data considered.\\

The Glasgow respiratory disease example gives insight into the model's performance for real data, which invariably contain more complex spatial structure than is present in idealised simulated data. Figure \ref{figure results} shows that the risk surface contains highly localised spatial structure, with some areas that are geographically adjacent having very different risks. However, this example does illustrate one potential extension that could be made to the model, namely that the boundaries could be forced to be closed. Enforcing this constraint is attractive because it would partition the study region into non-overlapping groups of areas that internally have similar risks, which would allow the spatial extent of high risk clusters to be more easily identified. However, as highlighted in the previous section, the spatial structure in real data is not that simple, and such an approach would only be appropriate if the goal of the analysis was cluster detection, rather than explaining the spatial pattern in the response. Another important area of future research is to extend the method adopted here so that the inherent uncertainty in $W$ is accounted for in the modelling process. Such as extension seems most likely using MCMC simulation, with $W$ being deterministically updated at every iteration. Finally, other extensions we will investigate include the detection of boundaries in multiple responses simultaneously, as well as adding a temporal dimension to the model.

\section*{Acknowledgements}
 The work of both authors was supported by the Economic and Social Research Council [RES-000-22-4256]. The data and shapefiles used in this study were provided by the  Scottish Government.

\section*{Supplementary material}
 R functions to implement the models developed in this paper together with a short tutorial on their use are available in the supplementary material accompanying this paper.

\bibliographystyle{chicago}
\bibliography{Lee}

\end{document}